\documentclass[superscriptaddress,aps,pra,twocolumn,showpacs,preprintnumbers,amsmath,amssymb,floatfix,groupedaddress]{revtex4}

\usepackage{graphicx}
\usepackage{dcolumn}
\usepackage{bm}

\begin{document}
\def\e{\mathcal{E}}

\title{Optimal light storage in atomic vapor}


\author{Nathaniel B. Phillips}
\affiliation{Department of Physics, College of William and Mary, Williamsburg,
Virginia 23185, USA}
\author{Alexey V. Gorshkov}
\affiliation{Department of Physics, Harvard University, Cambridge,
Massachusetts 02138, USA}
\author{Irina Novikova}
\affiliation{Department of Physics, College of William and Mary, Williamsburg,
Virginia 23185, USA}

\date{\today}

\begin{abstract}
We study procedures for the optimization of efficiency of light storage and
retrieval based on the dynamic form of electromagnetically induced transparency
(EIT) in warm Rb vapor. We present a detailed analysis of two recently
demonstrated optimization protocols: a time-reversal-based iteration procedure,
which finds the optimal input signal pulse shape for any given control field,
and a procedure based on the calculation of an optimal control field for any given
signal pulse shape. We verify that the two procedures are consistent with each
other, and that they both independently achieve the maximum memory efficiency for any given
optical depth. We observe good agreement with theoretical predictions
for moderate optical depths ($< 25$), while at higher optical depths the
experimental efficiency falls below the theoretically predicted values. We
identify possible effects responsible for this reduction in memory efficiency.

\end{abstract}


%
%

\maketitle

\section{Introduction}

The ability to store light pulses in matter and then retrieve them while preserving
their quantum state is an important step in the realization of quantum networks
and certain quantum cryptography protocols \cite{bouwmeester00,DLCZ}. Mapping
quantum states of light onto an ensemble of identical radiators (\emph{e.g.}, atoms, ions,
solid-state emitters, \emph{etc.}) offers a promising approach to the practical
realization of quantum
memory~\cite{fleischhauer,lukin03rmp,julsgaard04,kraus06}. Recent realizations
of storage and retrieval of single-photon wave
packets~\cite{eisaman05,kuzmich05, choi08}, coherent states~\cite{julsgaard04},
and squeezed vacuum pulses~\cite{appel,furusawa}
constitute an important  step in demonstrating the
potential of this method. However, the efficiency and fidelity of the storage
must be significantly improved before practical applications become
possible.

In 
this paper, we present a comprehensive analysis of two recently
demonstrated optimization protocols~\cite{novikovaPRLopt, novikova08prep}
that are based on a recent theoretical
proposal~\cite{gorshkovPRL,gorshkovPRA1,gorshkovPRA2,gorshkovPRA3,
gorshkovPRA4}. The first protocol iteratively optimizes the
input pulse shape for any given control field~\cite{novikovaPRLopt}, while the
second protocol uses optimal control fields calculated for any given input
pulse shape \cite{novikova08prep}. We experimentally
demonstrate their mutual consistency by showing that both protocols yield the
same optimal control-signal pairs and memory efficiencies. We also show that
for moderate optical depths ($\lesssim 25$), the experimental results presented here (as well as in Refs.\ \cite{novikovaPRLopt, novikova08prep}) are in excellent
agreement with a simple three-level theoretical model~\cite{gorshkovPRL,gorshkovPRA2} with no free parameters;
we discuss the details of the correspondence between the actual atomic system and this simple model. Lastly, we study the dependence of memory efficiency on the
optical depth. We show that for higher optical depths ($\gtrsim 25$), the experimental
efficiency falls below the theoretically predicted values; we discuss possible effects, such as spin-wave decay and four-wave mixing, that may limit
the experimentally observed memory efficiency.

%
\begin{figure}[!h]
\includegraphics[width=1.0\columnwidth]{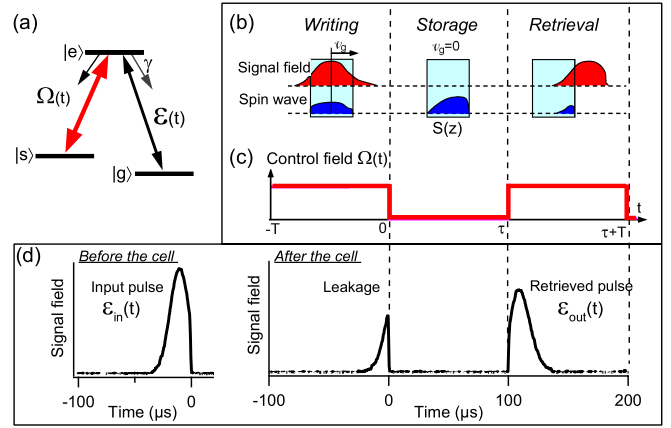}%
\caption {(Color online) (a) The three-level $\Lambda$ scheme used in
theoretical calculations. The schematic (b) and example control (c) and signal
(d) fields during light storage. At the writing stage ($t < 0$), an input
signal pulse $\e_\textrm{in}(t)$ propagates through the atomic medium with low
group velocity $v_\textrm{g}$ in the presence of a control field envelope
$\Omega(t)$. While compressed inside the cell, the pulse is mapped onto a
spin-wave $S(z)$ by turning the control field off at time $t=0$. After a
storage period $\tau$, the spin-wave is mapped back into an output signal pulse
$\e_\textrm{out}(t)$ using the retrieval control field envelope $\Omega(t)$ ($t
> \tau$).
\label{storagecartoon.fig}}
\end{figure}

The remainder of the paper is organized as follows: In Sec.\
\ref{sec:theory}, we briefly summarize the 
three-level theory
governing the two procedures for optimizing photon storage. In Sec.\ \ref{sec:exp}, we
describe our experimental system 
and discuss its correspondence to the three-level model. In Secs.\
\ref{sec:iter} and \ref{sec:contr}, we  present the results of experimental
studies of both optimization procedures and demonstrate their consistency. 
In Sec.\ \ref{sec:od}, we investigate the
dependence of memory efficiency on the optical depth of the medium. Finally, in
Sec.\ \ref{sec:conc}, we conclude with the summary of our results.

\section{Review of the theory\label{sec:theory}}

In this section, we briefly  review the necessary concepts  from the
theoretical work~\cite{gorshkovPRL, gorshkovPRA2} on which our experiments
rely.  We consider the propagation of a weak signal pulse with envelope $\e(t)$ and a
strong (classical) control field with a Rabi frequency envelope $\Omega(t)$
\cite{Omeganote} in a resonant $\Lambda$-type atomic medium under the conditions of electromagnetically induced transparency (EIT), as shown in Fig.~\ref{storagecartoon.fig}(a). The control field
creates a strong coupling between the signal field and a collective atomic spin
excitation (spin wave) \cite{lukin03rmp}. As a result, the initial pulse gets spatially compressed and slowed down inside the atomic ensemble. The group
velocity of the pulse is proportional to the control field intensity~\cite{fleischhauerRMP05}:
\begin{equation} \label{vg}
v_\textrm{g} \approx 2|\Omega|^2/(\alpha \gamma)\ll c,
\end{equation}
where 
$\gamma$ is the decay rate of the optical
polarization and $\alpha$ is the absorption coefficient (\emph{i.e.}, unsaturated
absorption per unit length),  so that $\alpha L$ is the optical depth of an
atomic medium of length $L$~\cite{dnote}.

Fig.~\ref{storagecartoon.fig}(b) illustrates schematically the three stages of
the light storage process (writing, storage, and retrieval), while Figs.\
\ref{storagecartoon.fig}(c) and \ref{storagecartoon.fig}(d) show control and
signal fields, respectively, during a typical experimental run. At the writing
stage, a signal pulse $\e_\textrm{in}(t)$ is mapped onto the collective spin
excitation $S(z)$ by adiabatically reducing the control field to zero. This
spin wave is then preserved for some storage time $\tau$ (storage stage), during which all optical fields are turned off. Finally, at the retrieval stage, the
signal field $\e_\textrm{out}(t)$ is retrieved by turning the control field
back on~\cite{lukin03rmp,fleischhauer}. In the ideal case, the retrieved signal
pulse is identical to the input pulse, provided the same constant control power is used at the writing and the retrieval stages. However, to realize this ideal storage,
two conditions must be met. On the one hand, the group velocity $v_\textrm{g}$
of the signal pulse inside the medium has to be low enough to spatially
compress the whole pulse into the length $L$ of the ensemble and avoid
``leaking'' the front edge of the pulse past the atoms. This requires
$\mathrm{T}v_\textrm{g} \ll L$, where $\mathrm{T}$ is the duration of the
incoming signal pulse. On the other hand, all spectral components of the
incoming pulse must fit inside the EIT transparency window to minimize
spontaneous emission losses $1/\mathrm{T} \ll \Delta \omega_\textrm{EIT} \simeq
\sqrt{\alpha L} v_\textrm{g}/L$ \cite{fleischhauer}. The simultaneous
satisfaction of both conditions is possible only at very high optical depth
$\alpha L \gg 1$ \cite{fleischhauer, gorshkovPRL}.

Experimental realization of very high optical depth in atomic ensembles requires high atomic density and/or large sample length. At high atomic
density, EIT performance can be degraded by competing 
processes, such as stimulated Raman scattering and four-wave mixing
\cite{lukinPRL97, lukinPRL98, haradaFWM, kangFWM, narducciFWM, agarwalFWM}.
Furthermore, spin-exchange collisions \cite{happer72}
and radiation trapping \cite{radtrapPRL,radtrapJMO,kleinSPIE08} may reduce
spin wave lifetime by orders of magnitude, limiting storage time and signal
pulse durations. In addition, achieving high optical depth in some experimental
arrangements may be challenging, such as in magneto-optical traps (see
\emph{e.g.},\ Refs.\ \cite{kuzmich05, choi08}). Therefore, it is crucial to be able to
maximize memory efficiency by balancing the absorptive and leakage losses at
moderately large $\alpha L$ via optimal shaping of control and/or signal temporal profiles.
To characterize our memory for light, we define memory
efficiency  $\eta$ as the probability of retrieving an incoming single photon
after storage, or, equivalently, as the energy ratio between initial and
retrieved signal pulses:
\begin{equation} \label{eta_defin}
\eta=\frac{\int_{\tau}^{\tau+\mathrm{T}}|\e_\textrm{out}(t)|^2dt}{\int_{-\mathrm{T}}^0|\e_\textrm{in}(t)|^2dt}.
\end{equation}
The goal of any optimization procedure then is to maximize $\eta$ under the
restrictions and limitations of a given system.

In the theoretical treatment of the problem, the propagation of a signal pulse
in an idealized three-level $\Lambda $ system, shown in
Fig.~\ref{storagecartoon.fig}(a), is described by three complex, dependent
variables, which are functions of time $t$ and position $z$~\cite{lukin03rmp,
fleischhauer, gorshkovPRA2}. These variables are the slowly-varying envelope
$\e$ of the signal field, the optical polarization $P$ of the
$|g\rangle-|e\rangle$ transition, and the spin coherence $S$. The equations of
motion for these variables are~\cite{lukin03rmp, fleischhauer, gorshkovPRA2}
\begin{eqnarray} \label{eq1}
(\partial_t + c \partial_z) \e(z,t)&=&i g \sqrt{N} P(z,t),
 \\
\partial_t P(z,t)&=&- \gamma P(z,t) + i g \sqrt{N} \e(z,t) + \nonumber \\
&& i \Omega(t-z/c) S(z,t),
\\ \label{eq3}
\partial_t S(z,t)&=&-\gamma_\textrm{s} S(z,t) + i \Omega(t-z/c) P(z,t),
\end{eqnarray}
where $g \sqrt{N} = \sqrt{\gamma \alpha c/2}$ is the coupling constant between the
atomic ensemble and the signal field, and
$\gamma$ and $\gamma_\textrm{s}$ are the polarization decay rates for the
transitions $|g\rangle-|e\rangle$ and $|g\rangle-|s\rangle$, respectively.
%
While, in general, Eqs.\ (\ref{eq1}-\ref{eq3}) cannot be fully solved analytically, they reveal several important properties of the optimization
process ~\cite{gorshkovPRL,gorshkovPRA2}. These properties are most evident in
the case when spin wave decay rate $\gamma_\textrm{s}$ is negligible during the
processes of writing and retrieval ($\gamma_\textrm{s} T \ll 1$), which will
hold for most of the discussion in the present paper, except for parts of Sec.\
\ref{sec:od}. In this case, the highest achievable memory efficiency depends
only on the optical depth $\alpha L$ and the mutual propagation direction of
the control fields during the writing and retrieval stages \cite{notecoprop}. For each optical
depth, there exists a unique spin wave, 
$S_{\mathrm{opt}}(z)$, which provides the maximum memory efficiency. Thus, the focus of the optimization process becomes identifying a
matching pair of \textit{writing} control and signal pulses 
that maps the signal pulse
onto this optimal spin wave. 
Note that no additional optimization is required with respect to 
the \emph{retrieval} control field, because the memory efficiency does not
depend on it, provided spin wave decay is negligible during retrieval
\cite{gorshkovPRL, gorshkovPRA2}.

In the present experiments and in Refs.\ \cite{novikovaPRLopt,novikova08prep},
the optimization procedures are tested using weak classical signal pulses
rather than quantum fields. Such experimental arrangements greatly improved the
experimental simplicity and the accuracy of data analysis. At the same time,
the linear equations of motion for classical and quantum signal pulses are
identical, which makes the presented results applicable to quantized signal
fields, such as, \emph{e.g.}, single photons. It is also important to note that the
original theoretical
work~\cite{gorshkovPRL,gorshkovPRA1,gorshkovPRA2,gorshkovPRA3, gorshkovPRA4}
considered a wide range of interaction processes for storing and retrieving
photon wave packets (\textit{e.g.,} EIT, far-off-resonant Raman, and spin echo
techniques) under a variety of conditions including ensembles enclosed in a
cavity \cite{gorshkovPRA1}, inhomogeneous broadening \cite{gorshkovPRA3}, and
high-bandwidth non-adiabatic storage ($1/T \sim \alpha L \gamma$)
\cite{gorshkovPRA4}. Since the proposed optimization procedures are, to a large
degree, common to all interaction schemes and conditions, our results  are
relevant to a wide range of experimental systems.

\section{Experimental arrangements \label{sec:exp}}

\begin{figure}
\includegraphics[width=1.0\columnwidth]{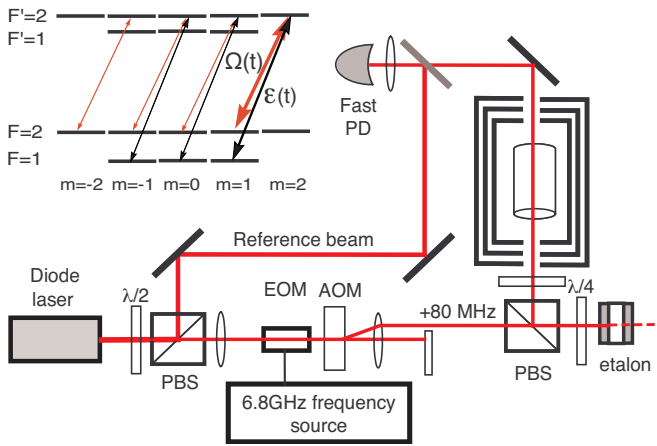}%
\caption {(Color online) Experimental apparatus (see text for abbreviations). \textit{Inset:}
Schematic of the ${}^{87}$Rb $\mathrm{D}_1$ line level structure and relevant
$\Lambda$ systems formed by control and signal fields. \label{setup.fig}}
\end{figure}

The schematic of the experimental apparatus is shown in Fig.\ \ref{setup.fig}.
We used an external cavity diode laser (ECDL) tuned near the $^{87}$Rb
$\mathrm{D}_1$ transition ($\lambda=795$~nm) with total available laser power
$\approx 45$~mW. After separating a fraction of original light for a reference
beam using a polarizing beam splitter (PBS), the main laser beam passed through
an electro-optical modulator (EOM), which modulated its phase at the frequency
of the ground-state hyperfine splitting of $^{87}$Rb
($\Delta_\mathrm{HF}=6.835$~GHz) and produced modulation sidebands separated by
that frequency. We tuned the zeroth order (carrier frequency) field to the $5^2
S_{1/2} F=2 \rightarrow 5^2 P_{1/2} F'=2$ transition. This field was used as
the control field during  light storage. The $+1$ modulation sideband played
the role of the signal field and was tuned to the $5^2 S_{1/2} F=1 \rightarrow
5^2 P_{1/2} F'=2$ transition.

To carry out the optimization procedure, we had to independently manipulate the
amplitudes of the control and the signal fields. We used an acousto-optical
modulator (AOM) to adjust the control field intensity. However, since all optical fields traversed the AOM, the intensities of all modulation comb fields were also changed.  Thus,  we accordingly adjusted the rf power at the EOM input (which controls the strength of the modulation sidebands) to compensate for any changes in the signal field amplitude caused by AOM modulation.

To minimize the effects of resonant four-wave mixing, we filtered out the other ($-1$)
first order modulation sideband (detuned by $\Delta_\mathrm{HF}$ to the red
from the carrier frequency field) by reflecting the modulation comb off of a
temperature-tunable Fabry-Perot etalon (FSR = 20~GHz, finesse $\approx 100$).
The etalon was tuned in resonance with this unwanted modulated sideband,
so that most of this field was transmitted.  At the same time, the control and signal
field frequencies were far from the etalon resonance, and were reflected back with no
losses. Such filtering allowed for suppression of the $-1$ modulated sideband
intensity by a factor of $\approx 10$. 

Typical peak control field and signal field powers 
were 18~mW and 50~$\mu$W, respectively.
The beam was weakly focused to $\approx 5$ mm diameter and circularly polarized with a
quarter-wave plate ($\lambda/4$).  A cylindrical Pyrex cell (length and
diameter were 75~mm and 22~mm, respectively) contained isotopically enriched
$^{87}$Rb and 30 Torr Ne buffer gas, so that the pressure broadened optical
transition linewidth was $2 \gamma = 2 \pi\times 290$ MHz~\cite{rotondaro97}.
The cell was mounted inside three-layer magnetic shielding to reduce stray
magnetic fields. The temperature of the cell was controllably varied between
$45^\circ$C and $75^\circ$C using a bifilar resistive heater wound around the
innermost magnetic shielding layer.

We used relatively short pulses, so that spin decoherence
had a negligible effect during writing and retrieval stages (except for parts of Sec.\ \ref{sec:od}) and only caused a
modest reduction of the efficiency 
during the storage time
$\propto \exp{(-2 \gamma_\textrm{s} \tau)}$. The Rb atom diffusion time out of
the laser beam ($\simeq 2$~ms) was long enough to avoid diffusion-related effects on
EIT dynamics~\cite{novikovaJMO05,xiaoPRL06}. We extracted the spin wave
decoherence time by measuring the reduction of the retrieved pulse energy as a
function of storage time and fitting it to an exponential decay. We found the
typical decay time to be $1/(2 \gamma_\textrm{s}) \simeq 500~\mu$s, most likely arising from small, uncompensated, remnant magnetic fields.

After the cell, the output laser fields were recombined with the reference beam
(at the unshifted laser frequency) at a fast photodetector, and the amplitude
of each field was analyzed using a microwave spectrum analyzer. Because of the
$80$~MHz frequency shift introduced by the AOM, the beatnote frequencies of the
$+1$ and $-1$ modulation sidebands with the reference beam differed by
$160$~MHz, which allowed for independent measurement of the amplitude of each
of these fields, as well as of the control field.

To conclude this section, we explain the direct correspondence between the
experimental system and the theory based on three-level atoms [Fig.\
\ref{storagecartoon.fig}(a)] that we reviewed in Sec.\ \ref{sec:theory}. The
goal is to use the structure of the $D_1$ line of $^{87}$Rb (see inset in
Fig.~\ref{setup.fig}) to identify the optical depth $\alpha L$ and the control
field Rabi frequency $\Omega$ for the effective three level system.
We first solve for the ground-state population distribution after control field optical
pumping of the Rb $D_1$ line, taking into account Doppler
broadening, pressure broadening, and collisional depolarization of the excited
state sublevels \cite{happer72}. We find the depolarization to be fast enough
(for $30$~Torr Ne, $\gamma_\textrm{depol} = 2 \pi \times 190$ MHz
\cite{rotondaro98})
 to ensure roughly equal populations in each of
$|F,m_F\rangle = |1,-1\rangle, |1,0\rangle, |1,1\rangle$, and $|2, 2\rangle$
ground state sublevels. Given this population distribution, we calculate the optical depth $\alpha L$ for the
signal field as a function of Rb number density. For example, we find that at
$60.5^\circ$C (Rb vapor density of $2.5\times 10^{11}~\mathrm{cm}^{-3}$) the
optical depth is $\alpha L=24.0$. Moreover, approximately  $60\%$ of this
optical depth comes from atomic population of  $|F,m_F\rangle=|1,1\rangle$ due
to the large corresponding Clebsch-Gordan coefficient. Thus, to calculate
$\Omega$, we use the dipole matrix element of the $|F=2,m_F=1\rangle
\rightarrow |F'=2,m_F=2\rangle$ transition.
Approximating  a transverse Gaussian laser beam profile with a uniform
cylindrical beam of diameter $5$ mm of the same power, we find, for example,
that for the control power of $16$ mW, $\Omega = 2 \pi\times
6.13$ MHz. Since the collisionally broadened optical transition linewidth ($2
\gamma = 2 \pi \times 290$ MHz) is comparable to the width of the Doppler
profile, the effects of Doppler broadening are negligible, making Eqs.
(\ref{eq1}-\ref{eq3}) directly applicable. We note that all the theoretical
modeling is done with no free parameters.

\section{Signal pulse optimization \label{sec:iter}}

One approach to the optimization of light storage is based on important time-reversal
properties of photon storage that hold even in the presence of irreversible
polarization decay~\cite{gorshkovPRA2}. In particular, for co-propagating \cite{notecoprop}
writing and retrieval control fields, the following is true under optimized
conditions (see Fig.~\ref{storagecartoon.fig}): if a signal pulse
$\e_\mathrm{in}(t)$ is mapped onto a spin wave using a particular control field
$\Omega(t)$ and retrieved after some storage time $\tau$ using the
time-reversed control field $\Omega(\mathrm{T}-t)$, the retrieved signal
pulse shape $\e_\mathrm{out}(t)$ is proportional to the time-reversed input
signal pulse $\e_\mathrm{in}(\mathrm{T}-t)$, but attenuated due to imperfect
memory efficiency. (Here and throughout the paper, control and signal envelopes are assumed to be real.)
This symmetry also gives rise to an experimentally
realizable iteration procedure, which, for any given writing control field,
determines the optimal incoming signal pulse shape.  This procedure
has been 
first demonstrated experimentally in Ref.~\cite{novikovaPRLopt}. The
present experiment was performed independently on a different (although
similar) experimental setup. Therefore, in order to use this procedure in Sec.\ \ref{sec:od} to study the dependence of memory efficiency on the optical depth, we verify in this section its successful performance in the present experimental setup. In addition, the implementation of iterative signal optimization in this experimental setup will allow us, in Sec.\ \ref{sec:contr}, to compare and verify the consistency of signal and control optimizations.

%
%
\begin{figure}
\includegraphics[width=1.0\columnwidth]{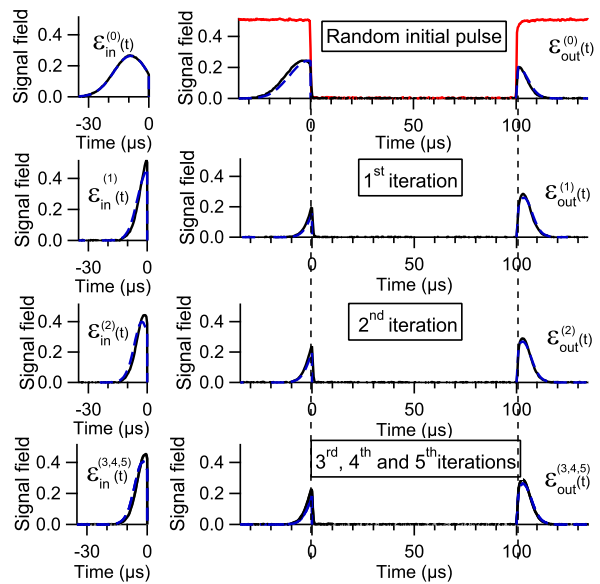}%
\caption {(Color online) Iterative signal pulse optimization. The experimental
data (solid black lines) is taken at $60.5~^\circ$C ($\alpha L=24$) using
$16$~mW constant control field during writing and retrieval (solid red line in
the top panel) with a $\tau = 100~\mu$s storage interval. Numerical simulations
are shown with blue dashed lines. \emph{Left}: Input pulses for each iteration.
\emph{Right}: Signal field after the cell, showing leakage of the initial pulse
for $t<0$ and the retrieved signal field $\e_\textrm{out}$ for $t>100~\mu$s.
Here and throughout the paper, all pulses are shown in the same scale, and all
input pulses are normalized to have the same area
$\int_{-\mathrm{T}}^0|\e_\textrm{in}(t)|^2dt = 1$, where $t$ is time in
$\mu$s.\label{iterations.fig}}
\end{figure}

The sequence of experimental steps for the iterative optimization procedure is
shown in Fig.~\ref{iterations.fig}. The plots show the control field and the measured and simulated  signal fields (solid red lines in the top panel, solid black lines, and dashed blue lines, respectively).
Before each iteration, we optically pumped all atoms into the state $|g\rangle$ by applying a strong
control field. We started the optimization sequence by sending an arbitrary signal pulse
$\e_{\mathrm{in}}^{(0)}(t)$ into the cell and storing it using a chosen control
field $\Omega(t)$. In the particular case shown in Fig.~\ref{iterations.fig},
the group velocity was too high, and most of the input pulse escaped the cell
before the control field was reduced to zero. However, a fraction of the pulse,
captured in the form of a spin wave, was stored for a  time period $\tau=100~
\mathrm{\mu s}$. We then retrieved the excitation using a time-reversed control field
$\Omega(t)=\Omega(\tau-t)$ and recorded the output pulse shape
$\e_{\mathrm{out}}^{(0)}(t)$. For the sample sequence shown, the control fields
at the writing and retrieval stages were constant and identical. This completes
the initial (zeroth) iteration step. 
The efficiency of light storage at this
step was generally low, and the shape of the output pulse was quite
different from the time-reverse of the initial pulse.
To create the input pulse $\e_{\mathrm{in}}^{(1)}(t)$ for the next iteration step, we digitally time-reversed the output $\e_{\mathrm{out}}^{(0)}(t)$ of the zeroth iteration
and renormalized it to compensate for energy
losses during the zeroth iteration: $\e_{\mathrm{in}}^{(1)}(t) \propto \e_{\mathrm{out}}^{(0)}(\tau-t)$. Then, these steps were repeated iteratively until the rescaled
output signal pulse became identical to the time-reversed profile of the input
pulse. As expected, the memory efficiency grew with each iteration and converged to $43 \pm 2\%$.


To verify that the obtained efficiency is indeed the maximum possible at this
optical depth and to confirm the validity of our interpretation of the results,
we compare the experimental data
to numerical simulations in Fig.\ \ref{iterations.fig}. Using the calculated
optical depth and the control Rabi frequency (see Sec.~\ref{sec:exp}), we solve
Eqs.\ (\ref{eq1}-\ref{eq3}) analytically in the adiabatic limit $T \alpha L
\gamma \gg 1$ \cite{gorshkovPRA2}, which holds throughout this paper.
There is a clear agreement between the
calculated and measured lineshapes and amplitudes of the signal pulses.
Also, theory and experiment converge to the optimal signal pulse shape in a
similar number of iteration steps (2-3), and the experimental efficiency ($43 \pm 2\%$) converged to a value close to the theoretical limit of $45\%$ (see below).

\begin{figure}
\includegraphics[width=1.0\columnwidth]{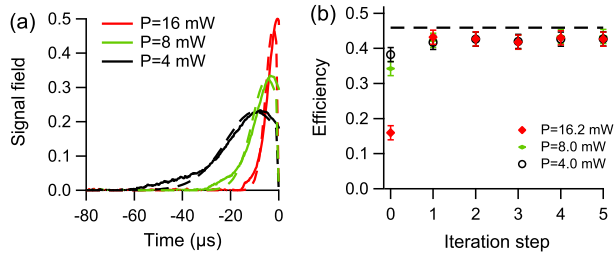}%
\caption { (Color online) (a) Experimental (solid) and theoretical (dashed) optimized signal pulses obtained after
five steps of the iteration procedure for three different powers of the constant control fields during writing and retrieval stages. (b) Corresponding memory
efficiencies determined for each iteration step. Theoretically predicted
optimal efficiency value is shown by the dashed line. The temperature of the cell
was $60.5~^\circ$C ($\alpha L=24$). \label{comp_eff_iter.fig}}
\end{figure}

As in our previous study~\cite{novikovaPRLopt}, we confirmed that the final
memory efficiency and the final signal pulse after a few iteration steps are
independent of the initial signal pulse $\e_{\mathrm{in}}^{(0)}(t)$. We also
confirmed that the optimization procedure yields the same memory efficiency for
different control fields. While  constant control fields of three different
powers yield different optimal signal pulses [Fig.\
\ref{comp_eff_iter.fig}(a)], the measured efficiency [Fig.\
\ref{comp_eff_iter.fig}(b)] converged after a few iteration steps to the same
value of $43 \pm 2\%$. With no spin wave decay, the highest achievable memory
efficiency for the optical depth $\alpha L = 24$ is $54
\%$~\cite{gorshkovPRA2}. Taking into account spin wave decay during the
$100~\mu$s storage time by a factor of $\exp[-100 \mu$s$/500 \mu$s$] = 0.82$,
the highest expected efficiency is $45\%$ [dashed line in Fig.\
\ref{comp_eff_iter.fig}(b)], which matches our experimental results reasonably
well.


\section{Control pulse optimization \label{sec:contr}}

\begin{figure*}
\includegraphics[width=1.5\columnwidth]{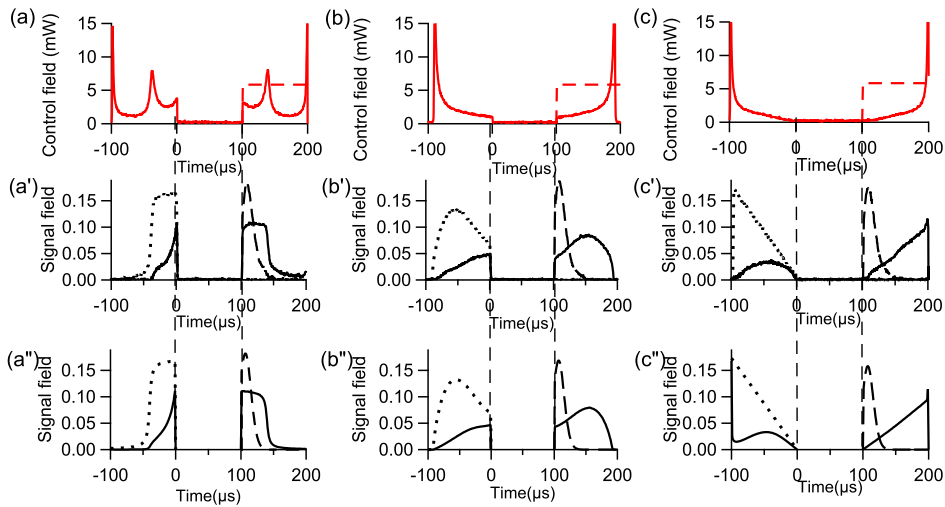}%
\caption {(Color online) Storage of three 
signal pulses (a$^{\prime}$, b$^{\prime}$, c$^{\prime}$) using calculated
optimal storage ($t < 0$) control fields (a), (b), (c).  Input signal pulse
shapes are shown in black dotted lines. The same graphs also show the leakage
of the pulses (solid black lines for $t < 0$) and retrieved signal pulses ($t > 100~
\mu$s) using flat control fields at the retrieval stage (dashed red lines), or
using time-reversed control fields (solid red lines). Graphs
(a$^{\prime\prime}$, b$^{\prime\prime}$, c$^{\prime\prime}$) show the results
of numerical calculations of (a$^{\prime}$, b$^{\prime}$, c$^{\prime}$).
The temperature of the cell was $60.5~^\circ$C ($\alpha
L=24$). 
\label{calculated.fig}}
\end{figure*}

The iterative optimization procedure described in the previous section has an
obvious advantage: the optimal signal pulse shape is found directly through
experimental measurements without any prior knowledge of the system parameters
(\emph{e.g.}, optical depth, control field Rabi frequency, various decoherence rates,
\emph{etc.}). However, in some situations, it is difficult or impossible to shape 
the input signal pulse (\emph{e.g.}, if it is generated by parametric down-conversion \cite{neergaard07}).  In these cases, the \emph{control field} temporal profile must be adjusted in order to optimally store and retrieve a given signal pulse.

To find the optimal writing control field for a given input pulse shape $\e_{\mathrm{in}}(t)$, we maximize $\eta$  [Eq.\ (\ref{eta_defin})] within the three-level model [Eqs.\ (\ref{eq1}-\ref{eq3})].
In this model, for a given optical depth $\alpha L$ and
a given retrieval direction (coinciding with the storage direction in the
present experiment \cite{notecoprop}), there exists an optimal spin wave
$S_{\mathrm{opt}}(z)$, which gives the maximum memory efficiency.
One way to
calculate the control field required to map the input pulse onto this optimal
spin wave is to first calculate  an artificial ``decayless'' spin wave
mode $s(z)$, 
which, like $S_{\mathrm{opt}}(z)$, depends only on the optical depth and not on the shape of the incoming pulse. This ``decayless" mode $s(z)$ hypothetically allows for unitary reversible storage of
an arbitrary signal pulse in a semi-infinite and polarization-decay-free atomic
ensemble, in which the group velocity of the pulse is still given by Eq. (\ref{vg}). The unitarity of the mapping
establishes a 1-to-1 correspondence between a given input signal pulse shape
$\e_{\mathrm{in}}(t)$ and an optimal writing control field that maps this input
pulse onto $s(z)$. The same control field maps this input pulse onto the true 
optimal spin wave $S_{\mathrm{opt}}(z)$, once polarization decay and the finite length of the
medium are taken into account. The details of this construction are described
in Ref.~\cite{gorshkovPRA2}.



As an example of control field optimization, we consider the storage of three
different initial pulse shapes, shown by dotted black lines in the middle row in
Fig.~\ref{calculated.fig}: a step with a rounded leading edge (a$^\prime$), a
segment of the sinc-function (b$^\prime$), and a descending ramp (c$^\prime$).
The top row (a,b,c) shows the corresponding calculated optimal writing ($t < 0$) control pulses.
Since the shape and power of the \emph{retrieval} control pulse do not affect the memory efficiency  \cite{gorshkovPRL, gorshkovPRA2}, we show, in the top row of Fig.~\ref{calculated.fig}, two retrieval control fields for each input
pulse: a flat control field (dashed) and the time-reverse of the writing
control (solid).
As expected, the flat control field (the same for all three inputs) results in
the same output pulse [dashed in (a$^\prime$, b$^\prime$, c$^\prime$)]
independent of the input signal pulse, because the excitation is stored in the
same optimal spin wave in each case. 
On the other hand, using the time-reversed writing control field for retrieval
yields output pulses that are time-reversed (and attenuated) copies of the corresponding input
pulses. This means that the time-reversal iterations of Sec.\ \ref{sec:iter}
starting with these control-signal pairs converge on the zeroth iteration,
which proves the consistency of the signal optimization of Sec.\ \ref{sec:iter}
with the control optimization of the present section. The experimental data also
agrees very well with numerical simulations [bottom row (a$^{\prime\prime}$,
b$^{\prime\prime}$, c$^{\prime\prime}$) in Fig.~\ref{calculated.fig}],
supporting the validity of our interpretation of the data.

\begin{figure*}
\includegraphics[width=1.7\columnwidth]{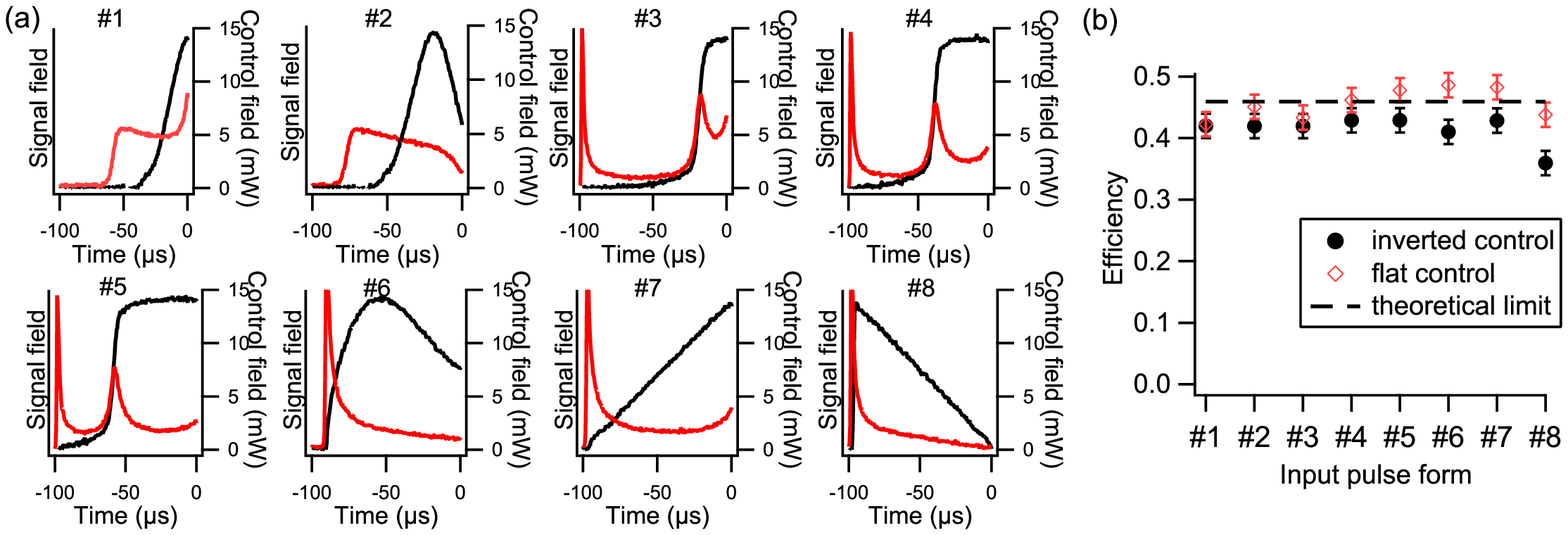}%
\caption {(Color online)  (a) Eight randomly selected signal pulse shapes
(black lines) and their corresponding optimal control fields (red lines). (b)
Memory efficiency for the eight signal pulse shapes using calculated optimized
control fields at the writing stage, and flat control fields (open red
diamonds) or inverted writing control fields (solid black circles) at the
retrieval stage. Theoretically predicted optimal memory efficiency is shown by
a dashed line. The temperature of the cell was $60.5~^\circ$C ($\alpha L=24$).
\label{comp_eff_calc.fig}}
\end{figure*}

To further test the effectiveness of the control optimization procedure, we
repeated the same measurements for eight different randomly selected pulse
shapes, shown as black lines in Fig.~\ref{comp_eff_calc.fig}(a). Pulses
$\#4$, $\#6$, and $\#8$ are the same as the input pulses (a$^\prime$),
(b$^\prime$), and (c$^\prime$) in Fig.\ \ref{calculated.fig}. For each of the
eight input pulses, we calculated the optimal writing control [red lines in
Fig.~\ref{comp_eff_calc.fig}(a)] and then measured the memory efficiency
[Fig.~\ref{comp_eff_calc.fig}(b)], retrieving with either a constant control pulse
or a time-reversed writing control pulse (open red diamonds and solid black
circles, respectively). The measured efficiencies are in good agreement with each other and
with the theoretically calculated maximum achievable memory efficiency  of
$45\%$ (horizontal dashed line) for the given optical depth.

By performing these experiments, we found that knowledge of accurate values
for the experimental parameters, such as optical depth or control field
intensity, is critical for calculations of the optimal control field. Even a few
percent deviation in their values caused measurable decreases in the output
signal pulse amplitude. In our experiment, effective optical depth and control
field Rabi frequency were computed accurately directly from measurable
experimental quantities with no free parameters. The accuracy of the parameters
was also verified by the excellent agreement of experimental and theoretical
results of iterative optimization in Sec.\ \ref{sec:iter}. We note that for
some other systems, the necessary experimental parameters may be difficult to
compute directly with high accuracy; in that case, they can be extracted from the
iteration procedure of Sec.\ \ref{sec:iter}.

\section{Dependence of memory efficiency on the optical depth\label{sec:od}}

\begin{figure*}
\includegraphics[width=1.4\columnwidth]{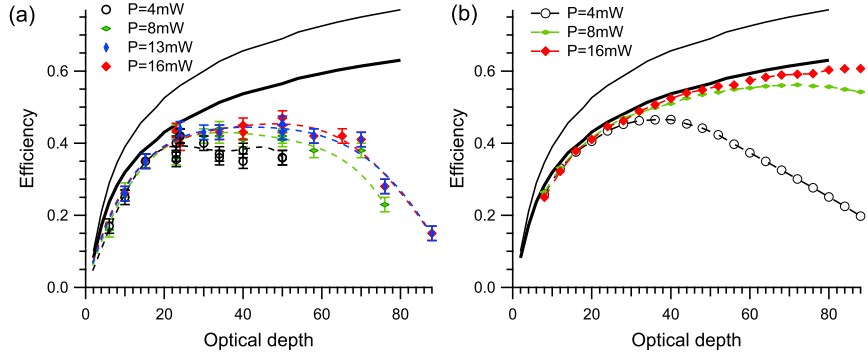}%
\caption {(Color online)  Memory efficiency as a function of optical depth
obtained by carrying out iterative signal optimization until convergence. (a)
At each optical depth, we considered constant control fields at four different
power levels (indicated on the graph) during writing and retrieval stages. Note
that many experimental data points overlap since the converged efficiencies are
often the same for different control fields. Dashed lines are to guide the eye.
Thin and thick black solid lines show the theoretically predicted maximum
efficiency assuming no spin-wave decay and assuming an efficiency reduction by
a factor of $0.82$ during the $100~\mu$s storage period, respectively. (b) Thin
and thick black lines are the same as in (a), while the three lines with
markers are calculated efficiencies for three different control fields
(indicated on the graph) assuming spin wave decay with a $500~\mu$s time constant
during all three stages of the storage process (writing, storage,
retrieval). \label{od_eff.fig}}
\end{figure*}

In the previous two sections, we verified at optical depth $\alpha L = 24$, the
consistency of the signal and control optimization methods and their agreement with
the three-level theory. In this section, we study the dependence of memory
efficiency on optical depth. To verify the theoretical prediction that the
optimal efficiency depends only on the optical depth of the sample, we repeated
the iterative signal optimization procedure (Sec.\ \ref{sec:iter}) for
several constant control field powers at different temperatures of the Rb
cell ranging from $45^\circ$C ($\alpha L=6$) to $77^\circ$C ($\alpha L=88$). In
Fig.~\ref{od_eff.fig}(a), we plot the  measured efficiencies (markers) along
with the maximum achievable efficiency predicated by the theory without spin
decay (thin black line) and with spin decay during the storage time (thick
black line). This graph allows us to make several important conclusions.

\begin{figure*}
\includegraphics[width=1.3\columnwidth]{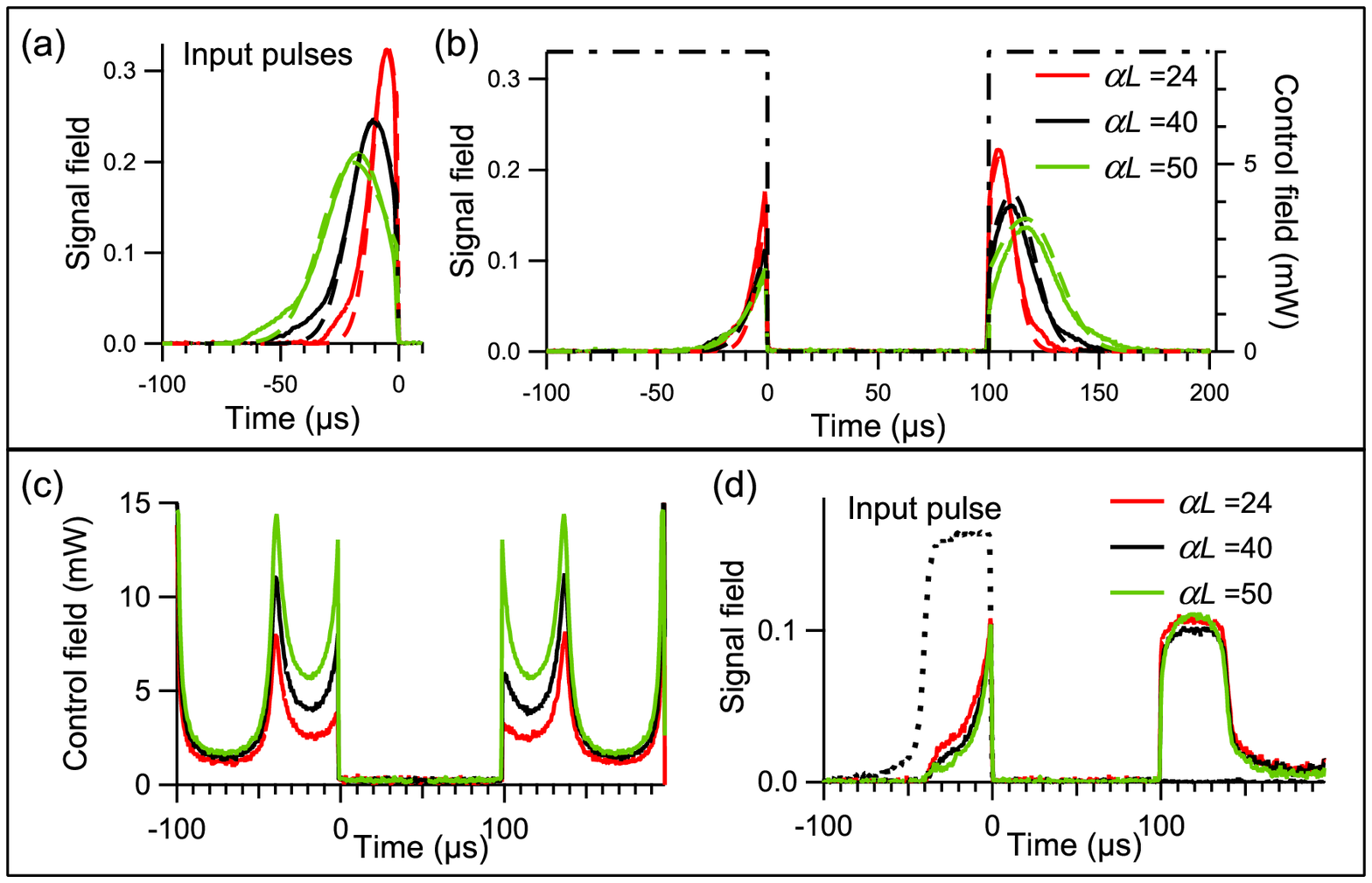}
\caption {(Color online) 
Results of the optimization procedures for different optical depths: $\alpha
L=24$ (red), 
$\alpha L=40$ (black), and 
$\alpha L=50$ (green). The top panel [(a) and (b)] shows storage and retrieval (b) of the optimized input signal pulses (a) obtained by running iterative optimization
until convergence for a constant control field of power $8$mW [dash-dotted line in (b)]. Solid lines
correspond to experimental results, while dashed lines show the results of
numerical simulations. In the bottom panel [(c) and (d)], (c) shows the calculated optimal writing control
fields ($t < 0$) for a step-like signal pulse [dotted line in (d)] and the time-reverses of these control fields used during retrieval ($t > 100~\mu$s), while (d) shows the resulting storage followed by
retrieval.
\label{od_pulses.fig}}
\end{figure*}

First of all, it demonstrates that for relatively low optical depths ($\alpha L
\le 25$), the optimized memory efficiency for different control fields is the
same, to within the experimental uncertainty, and approximately matches the
theoretical value (thick black line). This confirms that the optimization
procedure yields the maximum efficiency achievable for a given optical depth.
However, for $\alpha L
>20$, the efficiency obtained with the lowest control field power (black empty
circles) dropped below the efficiency obtained for higher control powers. As we
will now show, the most probable reason for such deviation is spin wave decay
during writing and retrieval.

As the optical depth increases, the duration of the optimal input pulse
increases as well, as shown in Fig.~\ref{od_pulses.fig}(a), following the
underlying decrease of group velocity: $T \sim L/v_\textrm{g} \propto \alpha L$
\cite{gorshkovPRA2}. Thus, above a certain value of $\alpha L$, the duration of
the optimal pulse for a given control field becomes comparable with the spin
wave lifetime, and the spin wave decoherence during storage and retrieval
stages can no longer be ignored. Further increase of the optical depth leads to
a reduction of retrieval efficiency, even though the iterative optimization
procedure is still valid \cite{gorshkovPRA2} and produces signal pulses that
are stored and retrieved with the highest efficiency possible for a given control
field and $\alpha L$. Fig.~\ref{od_eff.fig}(b) shows the calculated maximum
achievable efficiencies for different constant control powers as a function of
the optical depth, taking into account spin wave decay with a $500~\mu$s time
constant during all three stages of light storage. For each control field
power, the efficiency peaks at a certain optical depth, and then starts to
decrease as optical depth increases further. Since lower control powers require longer optimal input pulses $T \sim L/v_\textrm{g} \propto 1/|\Omega|^{2}$ [see
Fig.~\ref{comp_eff_iter.fig}(a)],  the corresponding efficiency reaches its peak at lower
optical depths. 
Thus, the problem of efficiency reduction posed by spin-wave
decay during writing and retrieval can be alleviated by using higher control
powers, and hence shorter optimal signal pulses. While this effect explains the
reduction of maximum memory efficiency attained with the lowest control power for $\alpha L > 20$
[Fig.\ \ref{od_eff.fig}(a)], other effects, discussed below, degrade the efficiency for
all other control powers for $\alpha L > 25$, as indicated by the divergence of experimental data
in Fig.\ \ref{od_eff.fig}(a) from the corresponding theoretical efficiencies in Fig.\ \ref{od_eff.fig}(b) (red and green lines).  Remarkably, at these optical depths, the iterative signal optimization procedure still yields efficiencies that grow monotonically at each iteration step for the three highest control powers.  This suggests that iterative signal optimization may still be yielding the optimum efficiency, although this optimum is lower than what the simple theoretical model predicts.

To further  test the applicability of our optimization procedures at higher
optical depths, we complemented the signal-pulse optimization [Fig.~\ref{od_pulses.fig}(a,b)] with the corresponding control field optimization [Fig.~\ref{od_pulses.fig}(c,d)]. We stored and retrieved 
input pulse
$\#4$ from Fig.~\ref{comp_eff_calc.fig}(a) using calculated optimal writing
control fields [$t<0$ in Fig.~\ref{od_pulses.fig}(c)] at different optical depths $\alpha L = 24$, $40$, and $50$.  
As expected, the
overall control power was higher at higher optical depths to keep the group
velocity unchanged: $L/T \sim v_g \propto \Omega^2/(\alpha L)$. For each
optical depth, we used a time-reversed writing control field to retrieve the stored
spin wave. 
This resulted in 
the output signal
pulse shape identical to the time-reversed (and attenuated) copy 
of the input pulse, as shown in
Fig.~\ref{od_pulses.fig}(d). 
Although the memory efficiency drops below the theoretical value at these high optical depths [$\alpha L = 50$ for the green lines in Fig.~\ref{od_pulses.fig}(c,d)], the results suggest that the calculated control field may still be optimal, since it yields the time-reverse of the input signal at the output.

\begin{figure}
\includegraphics[width=1.0\columnwidth]{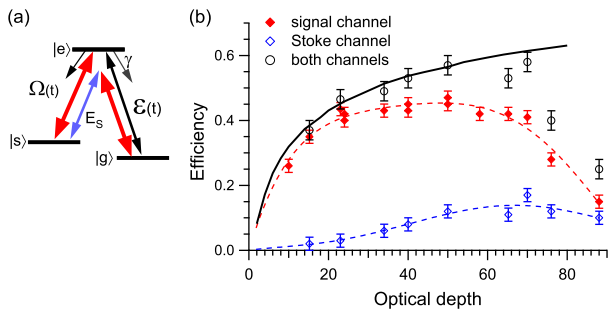}
\caption {(Color online) (a) Level diagram illustrating Stokes field
($\mathrm{E_S}$) generation due to
resonant four-wave mixing. 
(b) Memory efficiency for retrieval in the signal channel [same as the red filled diamonds in Fig.\ \ref{od_eff.fig}(a)], Stokes channel, and
the total for both channels. The efficiencies are obtained by carrying out
iterative optimization till convergence for constant writing and retrieval
control fields of $16$~mW power. Dashed lines are to guide the eye. The solid
line (same as the thick black line in Fig.\ \ref{od_eff.fig}) shows the theoretically
predicted maximum efficiency assuming an efficiency reduction by a factor of
$0.82$ during the $100~\mu$s storage period.
\label{SigStokesEff.fig}}
\end{figure}

To gain insight into what may limit the memory efficiency for 
$25<\alpha L<60$, we investigated the effect of resonant four-wave mixing.
Thus far, we have considered only the ground-state coherence created by the control
and signal fields in the one-photon resonant $\Lambda$ configuration [Fig.\
\ref{storagecartoon.fig}(a)]. However, the strong control field applied to the
ground state $|g\rangle$ can also generate an additional Stokes field
$\mathrm{E_S}$, as shown in Fig.~\ref{SigStokesEff.fig}(a). This process is
significantly enhanced in EIT media~\cite{lukinPRL97,lukinPRL98}. In
particular, it has been shown that a weak signal pulse traversing an atomic
ensemble with reduced group velocity generates a complimentary Stokes pulse
that travels alongside with a comparably low group
velocity~\cite{mikhailovJMO03,boyerPRL07}.

To determine the effect of resonant four-wave mixing on light storage, we first
carried out iterative signal optimization for a constant control field pulse of
$16$~mW power at different optical depths, but then
detected not only the signal field, but also the 
Stokes field, at the retrieval stage [see Fig.\ \ref{SigStokesEff.fig}(b)]. We see that
at low optical depths, the 
retrieved Stokes pulse [blue empty diamonds] 
is 
negligible compared to the output signal pulse [red filled diamonds, which are the same as the red filled diamonds in Fig.\ \ref{od_eff.fig}(a)]. However, at
$\alpha L \gtrsim 25$, 
the energy of the output pulse in the
Stokes channel becomes significant. While the energy of the retrieved signal
pulse stayed roughly unchanged for $25<\alpha L<60$, the energy of the output
Stokes pulse showed steady growth with increasing $\alpha L$. Moreover, the combined energy (black empty circles) of the two pulses
retrieved in the signal and Stokes channels added up to match well the
theoretically predicted highest achievable efficiency (solid black line). 
We will study elsewhere whether this match is
incidental and whether it can be harnessed for memory applications. For the
purposes of the present work, we simply conclude that the effects of four-wave
mixing become significant around the same value of $\alpha L$ ($\sim 25$) where experiment
starts deviating from theory. Therefore, four-wave mixing may be one of the
factors responsible for the low experimental efficiencies at high optical
depths.

For $\alpha L > 60$, iterative signal optimization still converges, but
efficiency does not grow monotonically at each iteration step, which clearly
indicates the breakdown of time-reversal-based optimization
\cite{gorshkovPRA2}. In addition, the final efficiency is significantly lower
than the theoretical value (Fig.\ \ref{od_eff.fig}). Many factors, other than
four-wave mixing, may be contributing to the breakdown of time-reversal-based
optimization and to the rapid decrease of memory efficiency at $\alpha L>60$.
First of all, the absorption of the control field at such high optical depths
is significant (measured to be $>50\%$). In that case, the reabsorption of
spontaneous radiation contributes appreciably to spin wave
decoherence~\cite{radtrapPRL,radtrapJMO} and can make the spin wave decay rate
$\gamma_\textrm{s}$ grow with $\alpha L$, reducing the light storage
efficiency~\cite{kleinSPIE08}. Spin-exchange collision rate \cite{happer72},
which destroys the spin-wave coherence, also becomes significant at high Rb
density, reducing spin wave lifetime even further.

\section{Conclusions \label{sec:conc}}

We have studied in detail two 
quantum memory optimization protocols in
warm Rb vapor and demonstrated their consistency for maximizing memory
efficiency. We have also observed good agreement between our experimental data
and theoretical predictions for relatively low optical depths ($< 25$), both
in terms of the highest memory efficiency and in terms of the optimized pulse shapes. At
higher optical depths, however, the experimental efficiency was lower than
predicted. We observed that resonant four-wave mixing processes became
important at these higher optical depths. We expect our studies to be of
importance for enhancing the performance of ensemble-based quantum memories for
light.

\section{Acknowledgments}

We are grateful to M. Klein, M. D. Lukin, A. S. S{\o}rensen, and Y. Xiao for useful discussions, and to J.
Goldfrank for assistance in experiments. This research was supported by
National Science Foundation, Jeffress Research grant J-847 and by the College
of William \& Mary.


\begin{thebibliography}{99}
\bibitem{Omeganote}
As in Refs.~\cite{fleischhauer,gorshkovPRL}, we define the Rabi frequency
$\Omega$ as $|\Omega|^2=\wp_{es}^2 I /(2 \hbar^2 \epsilon_0 c)$, where $I$ is
the control intensity.
\bibitem{notecoprop} The present experiment uses co-propagating writing and retrieval control fields, which corresponds to ``forward retrieval," using the terminology of Refs.\ \cite{gorshkovPRL, gorshkovPRA1, gorshkovPRA2, gorshkovPRA3, gorshkovPRA4}. Although backward retrieval is more efficient than forward retrieval for degenerate lower levels $|s\rangle$ and $|g\rangle$ of the $\Lambda$-system \cite{gorshkovPRL, gorshkovPRA2}, for the present experiment the hyperfine splitting between $|s\rangle$ and $|g\rangle$ makes forward retrieval more efficient \cite{gorshkovPRA2}.
\end{thebibliography}
\end{document}